\documentclass{iopart}

\usepackage{graphicx}  
\usepackage{latexsym}
\usepackage{amssymb}

\jl{6}        
\eqnobysec    

\def\beq{\begin{equation}}
\def\eeq{\end{equation}}

\begin{document}

\title[Schwarzschild black hole embedded in a dust field]
{Schwarzschild black hole embedded in a dust field: scattering of particles and drag force effects}

\author{
Donato Bini${}^*$ and
Andrea Geralico${}^{*,\dag}$
}

\address{${}^*$
Istituto per le Applicazioni del Calcolo ``M. Picone,'' CNR, 
I--00185 Rome, Italy}

\address{${}^\dag$
Astrophysical Observatory of Torino, INAF, via Osservatorio 20, I-10025 Pino Torinese (TO), Italy
}

\date{\today}

\begin{abstract}
A \lq\lq temporal analogue'' of the standard Poynting-Robertson effect is analyzed as induced by a dust of particles (instead of a gas of photons) surrounding a Schwarzschild black hole.
Test particles inside this cloud undergo acceleration effects due to the presence of a friction force, so that the fate of their evolution can be 
completely different from the corresponding geodesic motion.  
Typical situations are discussed of hyperbolic motion of particles scattered by the black hole in the presence of a dust filling the whole spacetime region outside the horizon as well as particles which free fall radially crossing a corona located at a certain distance from the horizon. 
The existence of equilibrium orbits may prevent particles from either falling into the hole or escaping to infinity.
\end{abstract}

\pacno{04.20.Cv}

\section{Introduction}

The spacetime region around compact objects is usually filled by accreting matter and radiation, which can be modeled by distributions of particles and photons endowed with suitable geometrical and thermodynamical properties, induced by the background gravitational field.
Let $T_{\mu\nu}$ be the energy-momentum tensor associated with the matter-energy field.
When the characteristic length scales of the thermodynamical quantities and their gradients maintain small with respect to the background curvature scale, the matter fluid/radiation field can be considered as a test fluid/field, in the sense that it does not back-react on the background metric.
Typical situations are represented by a photon flux outside a radiating source and a dust of particles surrounding a black hole, with associated energy-momentum tensors
\beq
\label{t_rad_matt}
T_{\rm (rad)}^{\mu\nu}=\Phi^2 k_{\rm g}^\mu k_{\rm g}^\nu\,,\qquad 
T_{\rm (dust)}^{\mu\nu}=\rho U_{\rm g}^\mu U_{\rm g}^\nu\,,
\eeq
respectively, where $k_{\rm g}^\mu$ and $U_{\rm g}^\mu$ denote the geodesics of the background, null and timelike respectively (fully known, also depending on several parameters like energy and angular momentum of the photons or the particles).
The flux factor $\Phi$ as well as the dust energy density $\rho$ are determined by the divergence-free condition $T^{\alpha\beta}{}_{; \beta}=0$ completely, if they depend in a simple way on the coordinates or in special situations like the one we are interested here of motion confined to the equatorial plane (of a metric which has this plane as a symmetry plane).

Poynting and Robertson \cite{PR} first investigated the effect of the radiation pressure of the light emitted by a star on the motion of small bodies orbiting it. They assumed a Thomson-type interaction, i.e., absorption and consequent re-emission of radiation by the particle, and modeled it by inserting a force term $f(U)$ into the equations of motion taken to be proportional to the 4-momentum density of radiation observed in the particle's rest frame by the (constant) effective interaction cross section $\lambda$ as follows
\beq
\label{force_U}
f(U)^\alpha=- \lambda P(U)^\alpha{}_\mu T^{\mu\nu}U_\nu\,,
\eeq
where $U$ denotes the particle 4-velocity, $P(U)$ projects orthogonally to $U$ as specified below, and $T^{\mu\nu}=T_{\rm (rad)}^{\mu\nu}$ is the stress-energy tensor associated with the radiation field given by  Eq. (\ref{t_rad_matt}).
The general relativistic formulation of the Poynting-Robertson effect has recently received much and novel attention to study the problem of scattering of test particles by a radiation field in different backgrounds of astrophysical interest \cite{abram,ML,Bini:2008vk,Oh:2010qn,Bini:2011zza,Bini:2014sua,Bini:2011zz,Bini:2014yca}, including the familiar Schwarzschild and Kerr spacetimes, radiating spacetimes (Vaidya), sources with non vanishing quadrupole moment (Erez-Rosen).

In the present paper we consider a temporal counterpart to the Poynting-Robertson effect, i.e., we study the features of particle's motion under the action of a friction force of the same form as Eq. (\ref{force_U}), but built with the stress-energy tensor $T^{\mu\nu}=T_{\rm (dust)}^{\mu\nu}$ associated with a distribution of collisionless dust around a Schwarzschild black hole (see Eq. (\ref{t_rad_matt})). 
A similar formulation has been adopted to study non-geodesic motion in the spacetime of self-gravitating fluids (also coupling the matter-energy distribution with the geometry) \cite{Bini:2012ff,Bini:2013es,Bini:2014lwz} as well as (test) perfect fluids describing either ordinary or exotic matter surrounding a non-rotating source \cite{Bini:2012zzd} from a cosmological perspective. 

We will consider first the case of an ingoing dust filling the whole spacetime outside the horizon of a Schwarzschild black hole, and study how the geodesic hyperbolic motion of particles scattered by the hole is modified under the action of the friction force due to the interaction with the surrounding dust.
We will then take the matter distribution to be confined to a corona located at a certain distance from the horizon, and determine the deviation of an initially (geodesic) radial trajectory when the particle enters the region occupied by the dust.   
In both cases we will find conditions for the existence of equilibrium solutions to the equations of motion which prevent particles from either falling into the hole or escaping to infinity.
This is a feature also shared by the standard Poynting-Robertson effect.  

We will use geometrical units and conventionally assume that greek indices run from $0$ to $3$, whereas latin indices run from $1$ to $3$.

\section{The equations of motion}

Consider a test particle moving on the equatorial plane of a Schwarzschild black hole with metric written in standard coordinates 
\begin{eqnarray}
ds^2&=&-N^2 dt^2+N^{-2} dr^2+r^2(d\theta^2 +\sin^2\theta d\phi^2)\,,
\end{eqnarray}
$N=\sqrt{1-\frac{2M}{r}}$ denoting the lapse function.
Let $U=U^\alpha \partial_\alpha$ be the particle's four velocity with $U^\alpha=dx^\alpha/d\tau\equiv x^\alpha{}'$ ($\tau$ being the proper time parameter, $\theta=\pi/2$ and $U^\theta=0$)
and $a(U)$ its four acceleration 
\begin{eqnarray}
\label{acc_comp}
a(U)&=&\left( U^t{}'+\frac{2M}{r^2 N^2}U^tU^r\right) \partial_t \nonumber\\
&+&\left(U^r{}'-rN^2U^\phi{}^2-\frac{M}{r^2N^2}U^r{}^2+\frac{MN^2}{r^2}U^t{}^2\right)\partial_r\nonumber\\
&+&\left(U^\phi{}'+\frac{2}{r} U^rU^\phi \right)\partial_\phi\,. 
\end{eqnarray}

Let us assume the Schwarzschild black hole be embedded in a dust (test) field, described by an energy-momentum tensor given by
\beq
T^{\alpha\beta}_{\rm(dust)}=\rho(r)U_{\rm g}^\alpha U_{\rm g}^\beta\,,
\eeq
with $U^\alpha_{\rm g}$ the unit tangent vector to an equatorial geodesic orbit 
\begin{eqnarray}\fl\quad
U^t_{\rm g}&=&\frac{E}{N^2}\,,\qquad
U^\phi_{\rm g}=\frac{L}{r^2}\,,\qquad
U^r_{\rm g}=\pm\left[E^2-N^2\left(1+\frac{L^2}{r^2}\right)\right]^{1/2}\,,
\end{eqnarray}
where $E$ and $L$ are the conserved energy and angular momentum per unit mass.  
It is customary to parametrize them by means of the eccentricity $e$ and the semi-latus rectum $p$ as
\begin{eqnarray}
\label{Emjm}
E^2 &=& \frac{(p-2)^2-4e^2 }{p(p-3-e^2)}\,,\qquad
L^2 = M^2 \frac{p^2}{(p-3-e^2)}\,.
\end{eqnarray}
The distance of minimum approach is given by 
\beq
\label{rmindef}
r_{\rm min}=\frac{Mp}{1+e}\,.
\eeq
Furthermore, we will limit our considerations to \lq\lq ingoing dust," i.e., $U_{\rm g}^r \le 0$. 
As stated above, the function $\rho$ follows from the conservation law $T^{\alpha\beta}{}_{; \beta}=0$ and, assuming for it a dependence on the radial variable only (besides $E$ and $L$), it turns out that it is uniquely determined as
\begin{eqnarray}
\label{rhosol}
\rho(r)&=& \frac{\rho_0}{r^2[ (E^2-N^2)  -L^2N^2/r^2 ]^{1/2}}=\frac{\rho_0}{r^2 |U_{\rm g}^r|}\,,
\end{eqnarray}
where we select $\rho_0$ (otherwise depending on an arbitrary integration constant) such that on the horizon  $\lim_{r\to 2M}\rho=1$, i.e., $\rho_0=4E M^2$.
As a consistency request, any length scale associated with the fluid should be small in comparison with the length scale associated with the hole, namely its mass $M$.

The force exerted by the test fluid on the particle (with 4-velocity $U$, different from $U_{\rm g}$)  is assumed to be
\begin{eqnarray}
\label{fdragdef}
f(U)^\alpha &=&-\lambda P(U)^\alpha{}_\mu T^{\mu\beta}_{\rm(dust)} U_\beta
= -\lambda \rho(r)  (U_{\rm g}\cdot  U)\, P(U)^\alpha{}_\mu U_{\rm g}^\mu\,,
\end{eqnarray}
where $P(U)^\alpha{}_\beta=\delta^\alpha_\beta+U^\alpha U_\beta$ projects orthogonally to $U$ (as it is necessary, since $U\cdot a(U)=0$ identically).
Decomposing $U_{\rm g}$ according to $U$, i.e., in the proper reference of the particle, yields
\beq
U_{\rm g}=\gamma(U_{\rm g},U)[U+\nu(U_{\rm g},U)]\,,
\eeq
with the spatial three-velocity of $U_{\rm g}$ with respect to $U$, $\nu(U_{\rm g},U)$, orthogonal to $U$, i.e., $U\cdot \nu(U_{\rm g},U)=0$.
We have then 
\beq
f(U)^\alpha =\lambda \rho \gamma(U_{\rm g},U)^2 \nu(U_{\rm g},U)^\alpha\,,
\eeq
with magnitude
\beq
|| f(U) ||=\lambda  \rho \gamma(U_{\rm g},U)^2 ||\nu(U_{\rm g},U)||\,.
\eeq
The motion of the particle with mass $m$ is described by the acceleration-equal-force equation
\beq
\label{eqm}
m a(U)=f(U)\,,
\eeq
which we will conveniently rewrite below in a frame adapted to the spacetime symmetries, i.e., in terms of Zero-Angular-Momentum-Observers (ZAMOs).

\section{Observer-dependent analysis of the motion}

A natural family of test observers in the Schwarzschild spacetime has its  world lines aligned with the coordinate time lines (ZAMOs) and unit tangent vector $n=N^{-1}\partial_t$. 
An adapted orthonormal spatial triad is 
\beq
e_{\hat r}=N\partial_r\,,\qquad
e_{\hat \theta}=\frac{1}{r}\partial_\theta\,,\qquad
e_{\hat \phi}=\frac{1}{r\sin\theta}\partial_\phi\,.
\eeq
The unit, timelike particle's 4-velocity can then be decomposed as \cite{Jantzen:1992rg,Bini:1997ea,Bini:1997eb}
\beq
U=\gamma(U,n)[n+\nu(U,n)]\,,\qquad \nu(U,n)=\nu(U,n)^{\hat a}e_{\hat a}\,,
\eeq
where $\gamma(U,n)=NU^t=(1-||\nu(U,n)||^2)^{-1/2}$, $\nu(U,n)^{\hat \theta}=0$ and
\beq
\nu(U,n)^{\hat r}=\frac{U^r}{N^2 U^t}\,,\quad
\nu(U,n)^{\hat \phi}=\frac{rU^\phi}{N U^t}\,,
\eeq
and $||\nu(U,n)||^2=[\nu(U,n)^{\hat r}]^2+[\nu(U,n)^{\hat \phi}]^2$.
It is also useful to split magnitude and unit vector of the spatial velocity as
\beq
\nu(U,n)=||\nu(U,n)||\hat \nu(U,n)
\equiv\nu(\sin \alpha e_{\hat r}+\cos \alpha e_{\hat \phi})\,,
\eeq
abbreviated as $\nu(U,n)= \nu \hat \nu$ to shorten notation.

Formally, the same expressions hold for $U_{\rm g}$ (with a label ${\rm g}$ added to the various quantities), so that it can be decomposed as
\beq
\label{Ugeo}
U_{\rm g}=\gamma_{\rm g}[n+\nu_{\rm g}(\sin \alpha_{\rm g} e_{\hat r}+\cos \alpha_{\rm g} e_{\hat \phi})]\,,
\eeq
with $\alpha_{\rm g}\in [-\pi/2,0]$ for ingoing dust. 
The conserved energy and angular momentum per unit mass write as
\beq
\label{ELgeo}
E=N\gamma_{\rm g}\,,\qquad
L=r\gamma_{\rm g}\nu_{\rm g}\cos \alpha_{\rm g}\,,
\eeq
implying that  
\beq
\label{geos}\fl\qquad
\nu_{\rm g}=\frac1{E}\sqrt{E^2-N^2}\,,\qquad
\gamma_{\rm g}=\frac{E}{N}\,,\qquad
\cos \alpha_{\rm g}=\frac{LN}{r\sqrt{E^2-N^2}}\,,
\eeq
and
\beq
\label{nug_cos}
\nu_{\rm g}\cos \alpha_{\rm g}=\frac{N}{r}\frac{L}{E}\,.
\eeq
We then find
\beq
U\cdot U_{\rm g}=-\gamma \gamma_{\rm g}{\mathcal V} \,,\qquad 
{\mathcal V}=1-\nu \nu_{\rm g}\cos \beta\,,\qquad
\beta=\alpha-\alpha_{\rm g}\,,  
\eeq
so that the frame components of the drag force (\ref{fdragdef}) are given by
\begin{eqnarray}
[f(U)]^{\hat t}&=&\lambda \rho \gamma \gamma_{\rm g}^2 {\mathcal V}  (1-\gamma^2 {\mathcal V})
=\nu\left[\sin\alpha[f(U)]^{\hat r}+\cos\alpha[f(U)]^{\hat \phi}\right]\,, \nonumber\\
{}[f(U)]^{\hat r}&=&\lambda \rho \gamma \gamma_{\rm g}^2 {\mathcal V}  (\nu_{\rm g}\sin \alpha_{\rm g} -\gamma^2 \nu {\mathcal V}\sin \alpha )\,,\nonumber\\
{}[f(U)]^{\hat \phi}&=&\lambda \rho \gamma \gamma_{\rm g}^2 {\mathcal V}  (\nu_{\rm g}\cos\alpha_{\rm g} -\gamma^2 \nu {\mathcal V}\cos \alpha )\,,
\end{eqnarray}
with 
\beq
\rho=\frac{4M^2}{r^2\nu_{\rm g}|\sin\alpha_{\rm g}|}\,.
\eeq
Note that $\rho$ diverges as the distance of minimum approach (\ref{rmindef}) is reached. Therefore, in this limit ($r\to r_{\rm min}$) the test field assumption for the dust distribution is no longer valid, as already stated, and back-reaction effects on the background metric cannot be neglected.

Similarly, the components of the acceleration (\ref{acc_comp}) become
\begin{eqnarray}\fl\quad
a(U)^t&=& \frac{\gamma^2 \nu}{2N^2 r}\left[2Nr \gamma\, \nu'+\sin\alpha (1-N^2)  \right]
=\frac{1}{N}[a(U)]^{\hat t}\,,\nonumber\\
\fl\quad
a(U)^r&=& \frac{1}{2r} \left[2Nr \sin\alpha \gamma^3\, \nu' +2rN \gamma \nu \cos\alpha\, \alpha' +\gamma^2 (1-N^2)-2N^2 \gamma^2\nu^2 \cos^2\alpha  \right]\nonumber\\
\fl\quad
&=& N [a(U)]^{\hat r}\,,\nonumber\\
\fl\quad
a(U)^\phi &=&  \frac{1}{r^2}\left[ r\gamma^3 \cos\alpha\, \nu' -\nu r \gamma \sin \alpha\, \alpha' +N\gamma^2 \nu^2 \sin \alpha \cos \alpha \right]
=\frac{1}{r}[a(U)]^{\hat \phi}\,.
\end{eqnarray}
Finally, the equations of motion (\ref{eqm}) write as
\begin{eqnarray}
\label{eqmoto}
\frac{d\nu}{d\tau} &=&\sigma\rho\frac{\gamma_{\rm g}^2{\mathcal V}}{\gamma^2\nu}(1-\gamma^2 {\mathcal V})-\frac{N\sin \alpha}{r\gamma}\nu_K^2\,, \nonumber\\
\frac{d\alpha}{d\tau }&=&\sigma\rho\frac{\gamma_{\rm g}^2\nu_{\rm g}{\mathcal V}}{\nu}\sin(\alpha_{\rm g}-\alpha)
+\frac{N\gamma\cos \alpha}{r\nu}(\nu^2-\nu_K^2) \,,
\end{eqnarray}
where $\sigma=\lambda/m$ and $\nu_K=\sqrt{M/(r-2M)}$ is the Keplerian velocity associated with circular geodesic motion.
These equations should then be completed by the temporal, radial and azimuthal evolution equations
\begin{eqnarray}
\label{eqmoto2}
\frac{dt}{d\tau}= \frac{\gamma}{N}\,,\qquad
\frac{dr}{d\tau}= \gamma N \nu\sin \alpha\,,\qquad
\frac{d\phi}{d\tau}= \frac{\gamma\nu }{r}\cos \alpha\,.
\end{eqnarray}

\begin{figure}
\centering
\[\begin{array}{cc}
\includegraphics[scale=0.3]{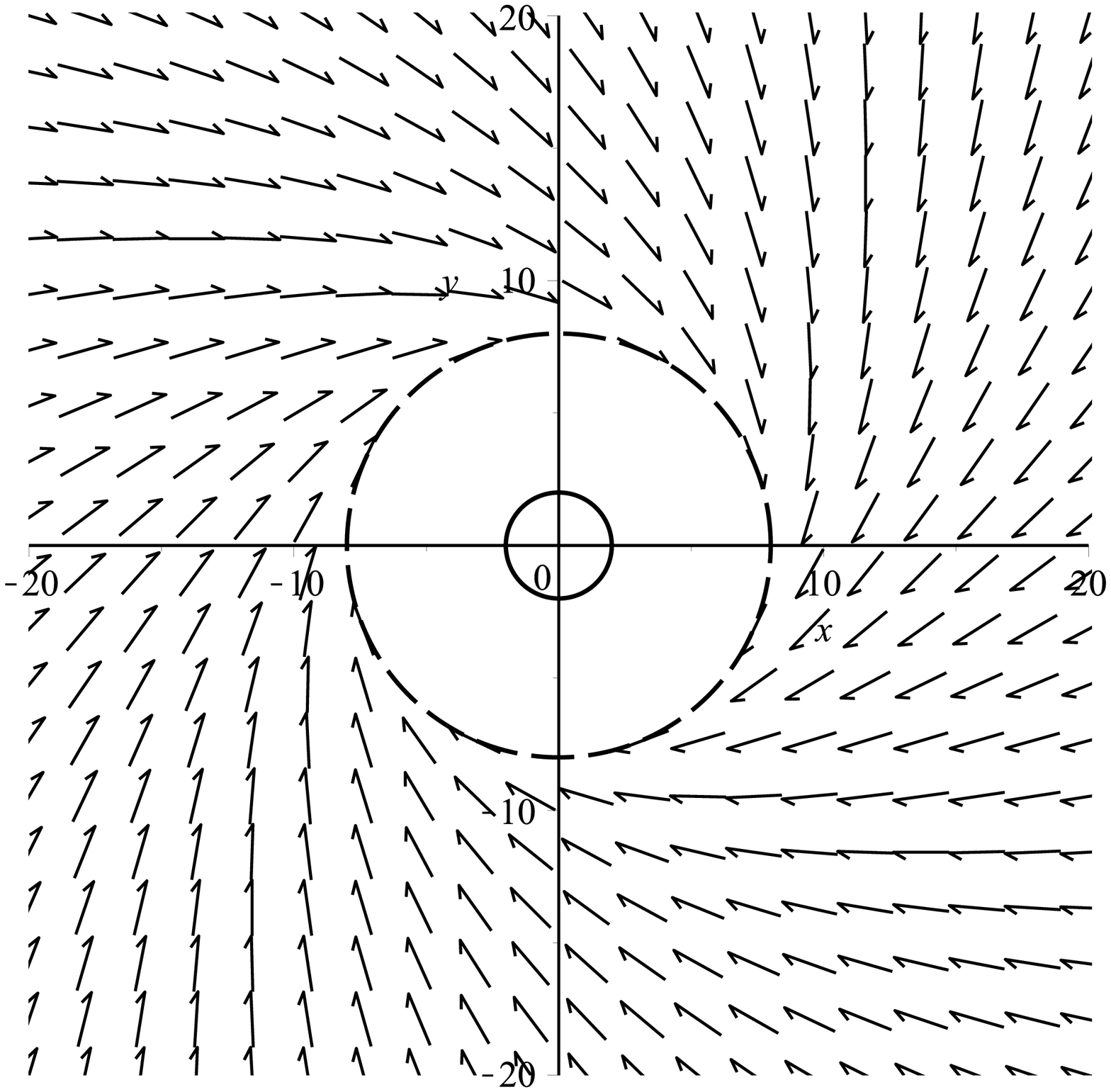}&\includegraphics[scale=0.3]{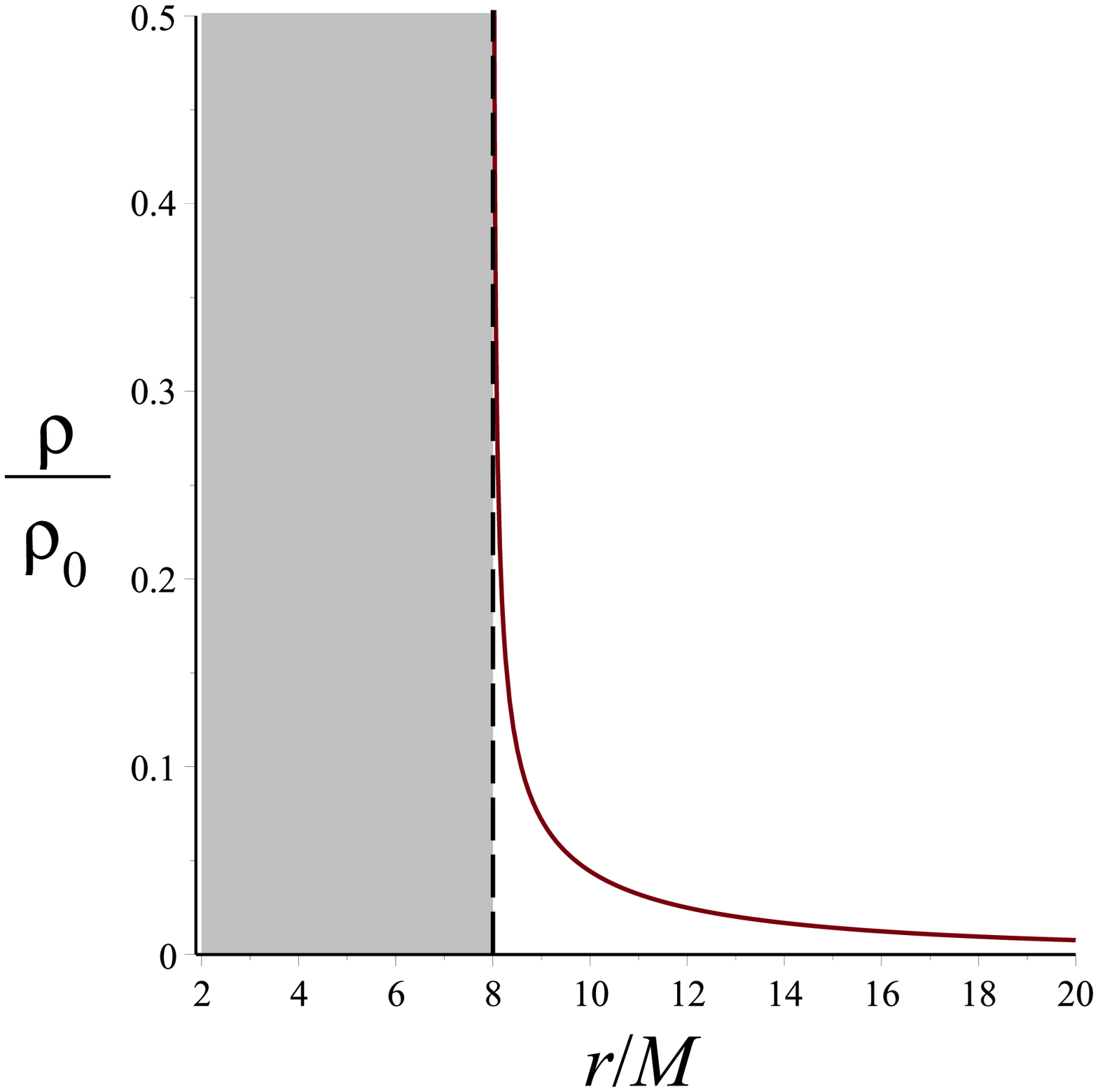}\\
(a)&(b)\\
\end{array}
\]
\caption{(a) The unit velocity direction field of ingoing dust is shown for $p=20$ and $e=1.5$ (so that the distance of minimum approach is $r=8M$).
(b) The density profile (in units of $\rho_0$) is shown for the same values of parameters as in (a). 
At $r=8M$ (dashed circle) the geodesic radial motion has an inversion point and the density of the dust diverges, so that the allowed region for the dust particles is $r>8M$ in this case.
}
\label{fig:1}
\end{figure}

Figure \ref{fig:1} shows the unit velocity direction field of ingoing dust for selected values of the parameters as well as the associated density profile. 
The semi-latus rectum and eccentricity are chosen as $p=20$ and $e=1.5$, implying that the particles of the dust cloud all move along eccentric geodesic with distance of minimum approach at $r_{\rm min}=8M$ (see Eq. (\ref{rmindef})). This is an inversion point for radial motion, so that the coordinate component of the geodesic four-velocity vanishes ($U^r_{\rm g}=0$) and correspondingly the density (\ref{rhosol}) diverges ($\rho\to\infty$).
Therefore, the allowed region for the distribution of dust particles in this case is $r>8M$.

                          
\subsection{Equilibrium solutions for ingoing dust}
\label{equilsols}

The system (\ref{eqmoto}) admits equilibrium solutions corresponding to $\alpha=\alpha_0=0$ and $\nu=\nu_0=$ const. at constant $r=r_0$, leading to the conditions
\begin{eqnarray}
\label{equil}
0 &=&-\sigma\rho\gamma_{\rm g}^2(\nu_0-\nu_{\rm g}\cos\alpha_{\rm g})(1-\nu_0\nu_{\rm g}\cos\alpha_{\rm g})\,, \nonumber\\
0&=&\sigma\rho\frac{\gamma_{\rm g}^2\nu_{\rm g}}{\nu_0}(1-\nu_0\nu_{\rm g}\cos\alpha_{\rm g})\sin\alpha_{\rm g}
+\frac{\gamma_0 N}{r\nu_0}(\nu_0^2-\nu_K^2)\,.
\end{eqnarray}
The previous equations admit the single (compatible) solution 
\beq
\label{nu0cond}
\nu_0=\nu_{\rm g}\cos\alpha_{\rm g}\,,
\eeq
or, by using Eq. (\ref{nug_cos}),
\beq
\nu_0=\frac{N}{r}\frac{L}{E}=Nu b\,,
\eeq
where $u=M/r$ and we have introduced the dimensionless quantity
\beq
b=\frac{L}{EM}\,.
\eeq
Inserting this solution into the second equation then implies 
\beq
0=\sigma\rho\gamma_{\rm g}^2\nu_{\rm g}\sin\alpha_{\rm g}
+\frac{N}{r}\gamma_0^3(\nu_0^2-\nu_K^2)\,,
\eeq
namely
\beq
\sigma=-\frac{rN}{4M^2}\frac{\gamma_0^3}{\gamma_{\rm g}^2}(\nu_0^2-\nu_K^2)\,{\rm sgn}(\alpha_{\rm g})\,.
\eeq
Using Eqs. (\ref{geos}) and (\ref{nug_cos}) the previous condition becomes
\beq
\label{equil_fin}
{\rm sgn}(\alpha_{\rm g})\,\tilde\sigma=N\gamma_0^3\left(1-\frac{\nu_0^2}{\nu_K^2}\right)=N\frac{1-N^4b^2u}{(1-N^2 b^2u^2 )^{3/2}}\,, 
\eeq
where $\tilde\sigma=4E^2M\sigma$, which is a relation between either $\nu_0$ or $b$ and $u$ (or equivalently $r$) for fixed values of $\tilde\sigma$.
It is worth noting that the equilibrium condition above does not depend explicitly on $E$, but only implicitly through the parameter $b$.
Furthermore, it is formally the same as the corresponding equilibrium condition for particles undergoing the standard Poynting-Robertson effect (see Eq. (3.1) of Ref. \cite{Bini:2011zza}), as discussed below. 

Finally, in the special case of dust particles in radial motion (i.e., $\alpha_{\rm g}=\pm \pi/2$) the equilibrium condition (\ref{nu0cond}) implies $\nu_0=0$ (i.e., $b=0$), so that Eq. (\ref{equil_fin}) becomes
\begin{eqnarray}
\label{equil2}
{\rm sgn}(\alpha_{\rm g})\tilde\sigma&=& N\,,
\end{eqnarray}
which admits solutions only for ${\rm sgn}(\alpha_{\rm g})=1$, i.e., outgoing dust, not considered here. 

\begin{figure}
\centering
\includegraphics[scale=0.35]{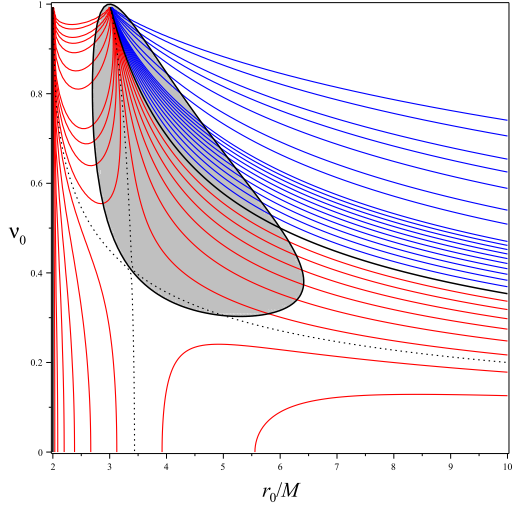}
\caption{The behavior of the linear velocity $\nu_0$ as a function of $r_0/M$ corresponding to equilibrium circular orbits is shown for equally spaced values of $\tilde\sigma$ from 0 to 1 at intervals of 0.1 and thereafter values of [1.5, 2, 3, 4, 5, 7.5, 10] in both cases of outgoing dust (red curves) and ingoing dust (blue curves).
The thick solid curve indicates the geodesic velocity $\nu_0=\nu_K$ for $\tilde\sigma=0$, distinguishing between the two cases.
The equilibrium solutions are stable outside the shaded region.
For ingoing dust, the equilibrium speed is always greater than the geodesic one ($\nu_0>\nu_K$), and increases at fixed $r_0$ for increasing values of $\tilde\sigma$ (so that curves are ordered from bottom to top). 
The outgoing dust case actually corresponds to a non-physical situation, and has been shown here only for completeness.
Its main features closely resemble those exhibited by the equilibrium orbits in the case of the standard Poynting-Robertson effect for an outgoing radiation field, including the change of the typical behavior as crossing a separatrix curve (dotted curve) as well as the existence of solutions very close to the source.  
}
\label{fig:2}
\end{figure}

Figure \ref{fig:2} shows the behavior of the equilibrium speed $\nu_0$ as a function of the equilibrium position $r_0/M$ for different values of $\tilde\sigma$.
The thick solid curve indicating the geodesic ($\tilde\sigma=0$) velocity $\nu_0=\nu_K$ distinguishes between the cases of outgoing dust (lower region, not discussed here) and ingoing dust (upper region).
The equilibrium solutions are stable outside the (superposed) shaded region. The boundary of the latter has been identified by solving the eingenvalue equation for the associated stability matrix (see \ref{appstab}).  
As mentioned above, the outgoing dust case is not explicitly discussed here, since it corresponds to a non-physical situation, but has been shown for completeness.
Furthermore, it is interesting to note the close similarity between the equilibrium solutions for ingoing/outgoing dust with those occurring in the case of an ingoing/outgoing radiation flux, leading to the standard Poynting-Robertson effect (see top panel of Fig. 1 in Ref. \cite{Bini:2011zza}). 

\begin{figure}
\centering
\includegraphics[scale=0.35]{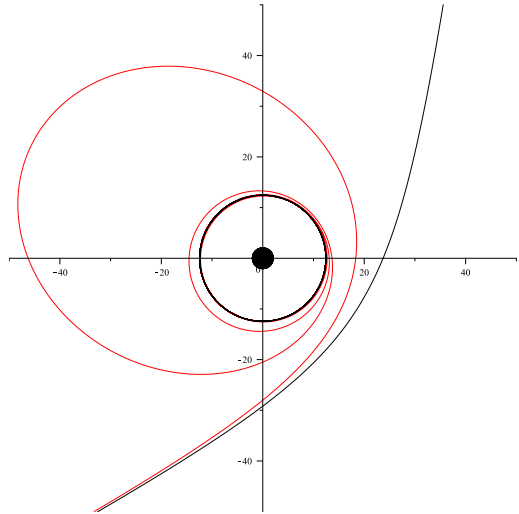}
\caption{Comparison between geodesic (black, $\tilde\sigma=0$) and accelerated orbit (red, $\tilde\sigma=0.5$).
In the latter case the trajectory ends up in a circular orbit at critical (equilibrium) radius $r_0/M\approx12.45$ (black circle) with speed $\nu_0\approx0.37$.
Initial conditions for both geodesic and accelerated orbits have been chosen as $r(0)/M=10^3$, $\phi(0)=-\pi/1.2$, $\alpha(0)=-\pi/2.04$ and $\nu(0)=0.3$.
The black disk represents the horizon.
}
\label{fig:3}
\end{figure}

An example of numerical integration of the orbits is shown in Fig. \ref{fig:3} for the same choice of semi-latus rectum and eccentricity as in Fig. \ref{fig:1} and a fixed value of the interaction parameter with the dust field.
The corresponding geodesic path is also shown for comparison.
The particle trajectory is dragged by the friction force, so that it cannot escape attraction, slows down spiraling around the hole and ends up in an equilibrium circular orbit.

                          
\subsection{Comparison with the standard Poynting-Robertson effect}

The physical mechanism underlying the standard Poynting-Robertson effect is driven by the presence of a photon flux which emanates from a radiating source, pushing particles {\it away} because of the associated radiation pressure.
Here we have considered, instead, a \lq\lq temporal analogue'' of the Poynting-Robertson effect due to a dust cloud surrounding a black hole and falling {\it towards} the hole itself, so dragging particles inward.
The main assumption in both cases is the presence of a (test) field superposed to the gravitational background spacetime, which induces acceleration effects on the motion of test particles. Such effects, however, have a different origin, namely either radiation pressure (pushing particles outward) or friction force (pushing particles inward).

In the standard Poynting-Robertson case the radiation pressure contrasts the gravitational attraction and gives rise to the existence of equilibrium solutions which have also attractive properties. The most natural outcome for a bunch of test particles  undergoing this effect is that it slows down and ends up into an equilibrium position at a certain radius. The latter can be either a fixed position in space (\lq\lq suspended orbit'') or corresponding to a circular orbit depending on the nature of spacetime (rotating or non-rotating) as well as on the properties of the photon flux (with or without angular momentum).

Interestingly, we have found the existence of equilibrium positions also in this \lq\lq temporal'' version. In particular, we have studied deviations from geodesic motion induced by this friction force in the case of particles scattered by the hole.
More precisely, a particle moving on such orbits has enough energy to escape attraction. The effect of the friction force is to cause a loss of both energy and angular momentum, so that the particle trajectory is deflected, starting then spiral motion around the hole until an equilibrium configuration is reached (see Fig. \ref{fig:3}).
In fact, from the point of view of the particle which experiences the drag, the friction force has both a radial component and a tangential component. It is just the latter which is responsible for contrasting the gravitational attraction, implying for the final state circular motion at a fixed radius and with fixed speed.
Summarizing, the standard Poynting-Robertson effect and its temporal counterpart show a number of close similarities, although associated with two different physical situations.

\section{Circularly rotating dust in a corona}

We can repeat the previous analysis by considering a different physical situation, namely that of a dust circularly rotating  in a corona around the hole.
In this case the congruence of timelike circular geodesics associated with the dust corresponds to Eqs. (\ref{Ugeo})--(\ref{geos}) with $\alpha_{\rm g}=0$, i.e.,
\beq
\label{Ugeocirc}
U_{\rm g}=\gamma_{\rm g}[n+\nu_{\rm g}e_{\hat \phi}]\,,
\eeq
with $\nu_{\rm g}=\sqrt{1-(N/E)^2}$.
Solving the conservation equations for the fluid gives $\rho=\rho_0=$ const., which we will set to unity without any loss of generality.
The equations of motion (\ref{eqmoto}) thus become
\begin{eqnarray}
\label{eqmotocirc}
\frac{d\nu}{d\tau} &=&\sigma\rho_0\frac{\gamma_{\rm g}^2{\mathcal V}}{\gamma^2\nu}(1-\gamma^2 {\mathcal V})-\frac{N\sin \alpha}{r\gamma}\nu_K^2\,, \nonumber\\
\frac{d\alpha}{d\tau }&=&-\sigma\rho_0\frac{\gamma_{\rm g}^2\nu_{\rm g}{\mathcal V}}{\nu}\sin\alpha
+\frac{N\gamma\cos \alpha}{r\nu}(\nu^2-\nu_K^2) \,,
\end{eqnarray}
where now ${\mathcal V}=1-\nu\nu_{\rm g}\cos\alpha$, still completed with Eqs. (\ref{eqmoto2}).
These equations are assumed to hold in the corona $r_{\rm (c,in)}\leq r\leq r_{\rm (c,out)}$, so that the full trajectory of the particle is made up of three pieces: geodesic for $2M\leq r\leq r_{\rm (c,in)}$ and $r\ge r_{\rm (c,out)}$, smoothly matched with the accelerated orbit in the spacetime region filled by the dust.

\begin{figure}
\centering
\[\begin{array}{cc}
\includegraphics[scale=0.3]{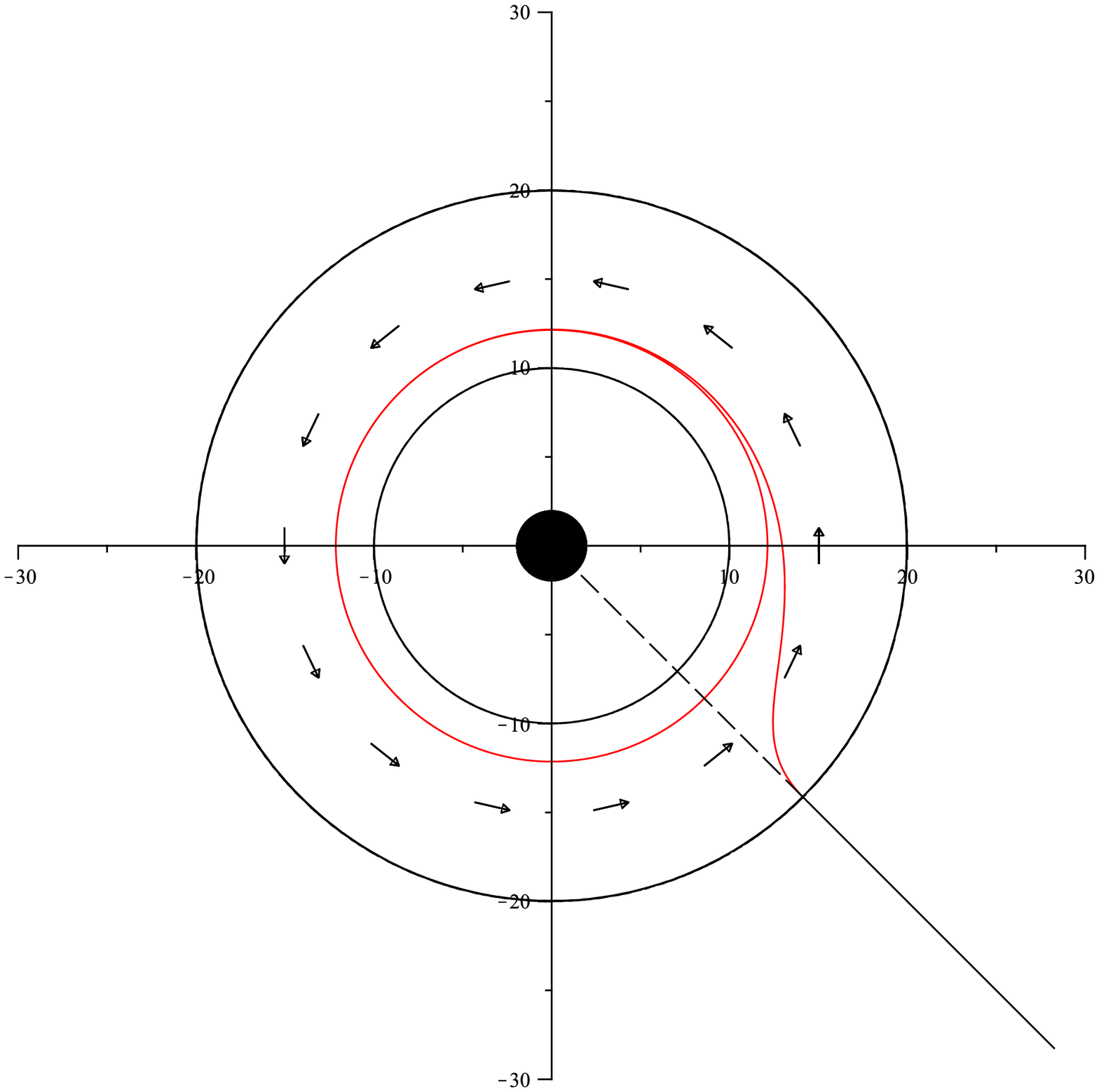}&\includegraphics[scale=0.3]{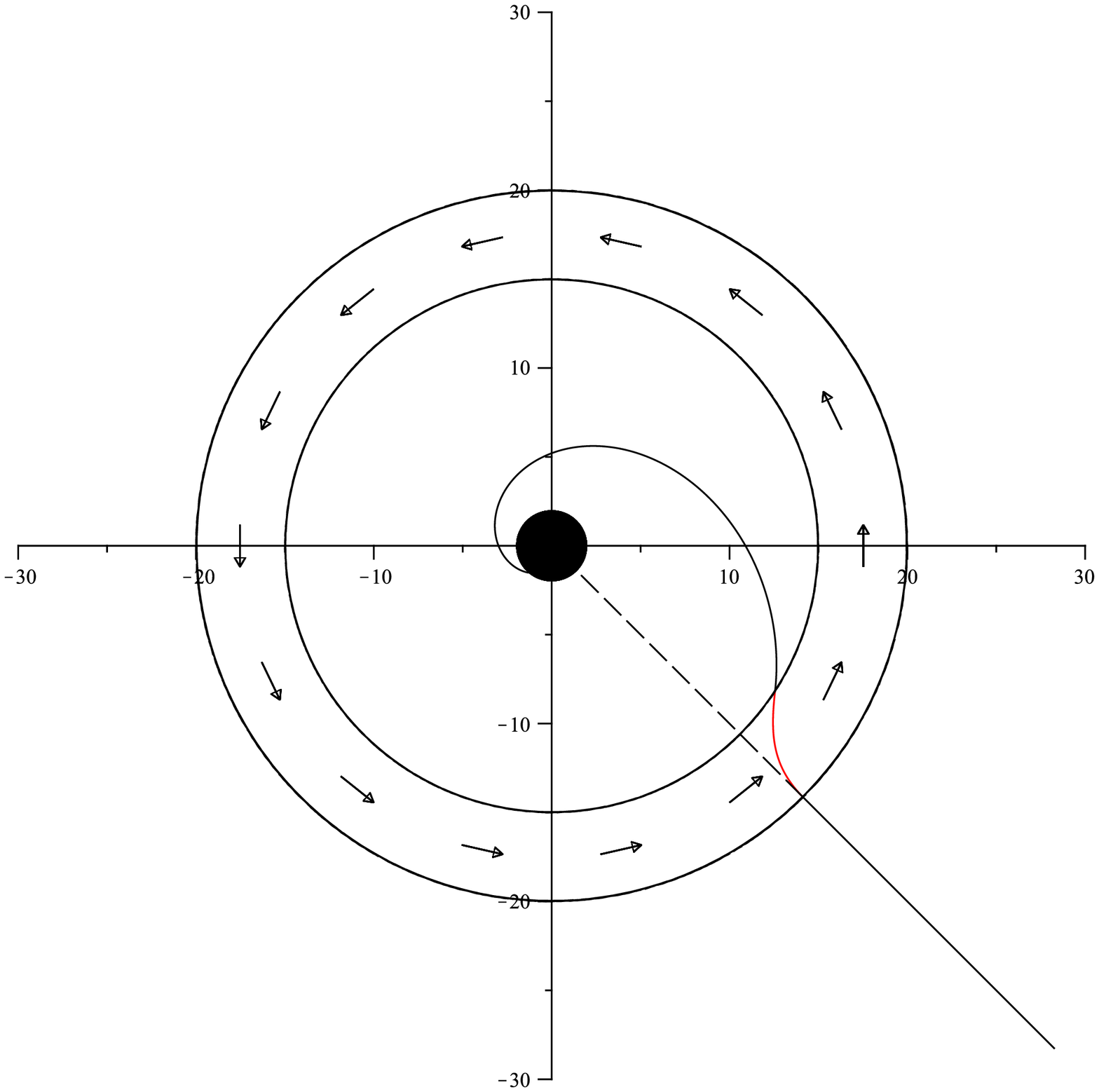}\\
(a)&(b)\\
\end{array}
\]
\caption{The trajectory of the accelerated particle is shown for $\tilde\sigma=0.05$ in two different situations. The common initial orbit is an ingoing ($\alpha(0)=-\pi/2$) radial geodesic starting at $r(0)/M=40$, $\phi(0)=-\pi/4$ with $\nu(0)=\sqrt{2M/r(0)}=1/\sqrt{20}$. The (counter-rotating) dust is located in a corona between $r_{\rm (c,out)}/M=20$ and (a) $r_{\rm (c,in)}/M=10$ and (b) $r_{\rm (c,in)}/M=15$.
In the former case there exists an equilibrium solution at $r_0/M\approx12.14$ with $\nu_0=\nu_K\approx0.31$ inside the corona, which acts as an attractor.
In the second case, instead, the orbit does not meet any equilibrium solution, and it ends up into the hole. 
A schematic representation of the velocity field of the dust is superposed to the plots.
}
\label{fig:4}
\end{figure}

Equilibrium solutions $\nu=\nu_0$, $\alpha=0$, $r=r_0$ exist also inside the corona, with the peculiar feature of being independent of $\sigma$ and corresponding to the geodesic value $\nu_0=\nu_K$ of the linear velocity at the equilibrium radius.
These features are illustrated in Fig. \ref{fig:4} in the two possible situations of a particle reaching the equilibrium position inside the corona and a particle falling into the hole.

                          
\section{Concluding remarks}

We have investigated a \lq\lq temporal analogue'' of the Poynting-Robertson effect as induced by a gas of dust particles surrounding a Schwarzschild black hole, instead of a gas of photons outside a radiating source as in the case of the standard Poynting-Robertson effect.
This is an interesting situation, since we expect that close to a black hole in special configurations (like accretion disks) a distribution of matter can be stored which may act as a drag force on particles moving there around.
The dust field is made of collisionless particles which move along geodesic paths on the equatorial plane with given energy and angular momentum. 

We have considered two different situations of an ingoing dust either filling the whole spacetime outside the black hole horizon or confined to a corona with fixed thickness located at a certain distance from the hole.
We have then explored the deviations from geodesic motion due to the friction force, whose main effect is to cause a loss of both energy and angular momentum, so that the particle trajectory gets deflected with respect to the corresponding geodesic. 
For instance, a particle moving on a scattering orbit may either escape to infinity (if it has enough energy and angular momentum) or slow down being definitely attracted to a certain equilibrium radius with a linear velocity greater than the geodesic speed, depending on the interaction strength of the dust field.
Equilibrium circular orbits also exist in the case of a matter distribution confined to a corona, but with linear velocity always equal to the geodesic speed. 

The existence of equilibrium configurations is a feature also shared by the standard Poynting-Robertson effect.  
However, despite the formal analogies, the physical processes underlying these effects are very different, so that a more realistic description of the interaction between particles and surrounding field is needed to develop a more accurate model for each effect.  
In particular, we have used the same force (\ref{force_U}) for both cases.
Different choices of the friction force may change completely the features of motion. For instance, one can consider a matter-like force field of viscous origin, i.e., a friction force whose components are proportional to those of the 4-velocity of the particle itself \cite{Gair:2010iv}.
In this case the equations of motion do not admit any equilibrium solution, as we will discuss in a forthcoming work.

\appendix

\section{Stability of the equilibrium solutions for ingoing dust}
\label{appstab}

The equilibrium solutions for ingoing dust discussed in Section \ref{equilsols} can be analyzed for their stability properties under small first order linear perturbations. 
Let 
\beq
r=r_0\,, \quad 
\phi=\phi_0(\tau)\,, \quad
\nu=\nu_0\not=0\,, \quad
\alpha=0
\eeq
be the parametric equation of the circular equilibrium orbit, or symbolically $X^\beta=X_0^\beta$ ($\beta=r,\phi,\nu,\alpha$). 
Consider the linear perturbations of this solution $X^\alpha=X_0^\alpha+X_1^\alpha$, namely
\beq
\fl\qquad
r=r_0+r_1(\tau)\ , \quad 
\phi=\phi_0(\tau)+\phi_1(\tau), \quad
\nu=\nu_0+\nu_1(\tau)\ , \quad
\alpha=\alpha_1(\tau)\,,
\eeq
which leads to the following linear system of constant coefficient homogeneous linear differential equations
\beq
\frac{dX_1^\alpha}{d\tau}=C^\alpha{}_\beta X_1^\beta\,,
\eeq
with nonzero coefficients 
\begin{eqnarray}\fl
C^1{}_{4}&=&\gamma_0\nu_0N_0\,,\qquad
C^2{}_{1} = -\frac{\gamma_0\nu_0}{r_0^2}\,,\qquad
C^2{}_{3} = \frac{\gamma_0^3}{r_0}\,,\nonumber\\
\fl
C^3{}_{1}&=& \frac{EN_0\gamma_0^2\nu_0}{\gamma_K^2r_0^2\sqrt{E^2-(N_0\gamma_0)^2}}(\nu_0^2-\nu_K^2)
\,,\qquad
C^3{}_{3} = \frac{EN_0\gamma_0^2}{r_0\sqrt{E^2-(N_0\gamma_0)^2}}(\nu_0^2-\nu_K^2)\,, \nonumber\\
\fl
C^3{}_{4}&=&-\frac{N_0\gamma_0\nu_0^2}{\gamma_K^2r_0}\,,\qquad
C^4{}_{1} = -\frac{N_0\gamma_0^3}{\nu_0r_0^2}[\nu_K^4+2\nu_K^2\nu_0^2(\nu_0^2-2)+\nu_0^2(2\nu_0^2-1)]
\,, \nonumber\\
\fl
C^4{}_{3}&=&\frac{2N_0\gamma_0^3}{\gamma_K^2r_0}\,,\qquad
C^4{}_{4} = \frac{N_0\gamma_0^2}{r_0E}\frac{2E^2-(N_0\gamma_0)^2}{\sqrt{E^2-(N_0\gamma_0)^2}}(\nu_0^2-\nu_K^2)\,.
\end{eqnarray} 
The associated eigenvalue equation is 
\beq
\lambda[\lambda^3+c_2\lambda^2+c_1\lambda+c_0]=0\,,
\eeq
where
\begin{eqnarray}\fl\qquad
c_0&=&-\frac{EN_0^3\gamma_0^6(\nu_0^2-\nu_K^2)}{r_0^3\sqrt{E^2-(N_0\gamma_0)^2}}[2(1+\nu_K^2)\nu_0^4+(1-8\nu_K^2+2\nu_K^4)\nu_0^2+\nu_K^4]\,,\nonumber\\
\fl\qquad
c_1&=&-\frac{N_0^4\gamma_0^6}{r_0^2[E^2-(N_0\gamma_0)^2]}\left\{
(3+2\nu_K^2)\nu_0^4+(1-10\nu_K^2+2\nu_K^4)\nu_0^2+2\nu_K^4\right.\nonumber\\
\fl\qquad
&&\left.
-\frac{E^2}{(N_0\gamma_0)^2}[2(2+\nu_K^2)\nu_0^4+(1-12\nu_K^2+2\nu_K^4)\nu_0^2+3\nu_K^4]
\right\}\,,\nonumber\\
\fl\qquad
c_2&=&-\frac{N_0\gamma_0^2}{r_0E\sqrt{E^2-(N_0\gamma_0)^2}}[3E^2-(N_0\gamma_0)^2](\nu_0^2-\nu_K^2)\,.
\end{eqnarray} 
The corresponding eigenvalues are then easily computed.

\section*{Aknowledgements}
D.B. thanks the Italian INFN (Naples) and  ICRANet for partial support.

\end{document}